\documentclass[superscriptaddress,prl,twocolumn]{revtex4} 

\usepackage{comment}
\usepackage{graphicx}
\usepackage{float}
\usepackage{epstopdf}
\usepackage{color}
\usepackage{amsmath}
\usepackage{amssymb}
\usepackage{braket}
\usepackage{chngcntr}
\renewcommand{\figurename}{Fig.}
\newcommand{\pare}[1]{\left(#1\right)}   
\newcommand{\avg}[1]{\left\langle #1\right\rangle}        

\newcommand{\equa}[2]{\begin{equation}\label{#1}\begin{aligned}#2\end{aligned}\end{equation}}

\newcommand{\kb}{k_\text B}             

\newcommand{\qf}{Q_0}                                     

\newcommand*\diff{\mathop{}\!\mathrm{d}}

\newcommand{\co}[1]{\cos{\left(#1\right)}}
\newcommand{\si}[1]{\sin{\left(#1\right)}}

\newcommand{\om}{\Omega_0}				
\newcommand{\fb}{F_\mathrm{fb}} 		
\newcommand{\gafb}{\Gamma_\mathrm{fb}}	
\newcommand{\xth}{x_\mathrm{th}}		

\begin{document}

\title{Thermodynamics of continuous non-Markovian feedback control}

\author{Maxime Debiossac}
\email{maxime.debiossac@univie.ac.at}
\affiliation{University of Vienna, Faculty of Physics, VCQ, Boltzmanngasse 5, A-1090 Vienna, Austria}
\affiliation{These authors contributed equally to this work}

\author{David Grass}
\affiliation{University of Vienna, Faculty of Physics, VCQ, Boltzmanngasse 5, A-1090 Vienna, Austria}
\affiliation{These authors contributed equally to this work}

\author{Jose Joaquin Alonso}
\affiliation{Department of Physics, Friedrich-Alexander-Universit\"at Erlangen-N\"urnberg, D-91058 Erlangen, Germany}

\author{Eric Lutz}
\affiliation{Institute for Theoretical Physics I, University of Stuttgart, D-70550 Stuttgart, Germany}

\author{Nikolai Kiesel}
\affiliation{University of Vienna, Faculty of Physics, VCQ, Boltzmanngasse 5, A-1090 Vienna, Austria}

\begin{abstract} 
\noindent {\bf \large{\textsf{Abstract}}}\\
Feedback control mechanisms are ubiquitous in science and technology, and play an essential role in regulating physical, biological and engineering systems. The standard second law of thermodynamics does not hold in the presence of measurement and feedback. Most studies so far have extended the second law for discrete, Markovian feedback protocols; however, non-Markovian feedback is omnipresent in  processes where  the control signal is applied with a non-negligible delay. 
Here, we experimentally investigate the thermodynamics of continuous, time-delayed feedback control using the motion of an optically levitated, underdamped microparticle. We test the validity of a generalized second law which bounds the energy extracted from the system and study the breakdown of feedback cooling for very large time delays.
\end{abstract}
\maketitle

\noindent {\bf \large{\textsf{Introduction}}}\\ 
The second law of thermodynamics is of fundamental and practical importance \cite{cen01}. On the one hand, it allows one to predict which transformations are  possible in Nature. On the other hand, it offers a  method for determining the efficiency of a given process by comparing it to its ideal, reversible limit. According to the standard formulation of the second law, no work can be cyclically extracted from a system coupled to a single reservoir at temperature $T_0$, that is, the power output has to be negative, $\dot{\mathcal{W}}_{\text{ext}}\leq0$ \cite{cen01}. However, in the presence of measurement and feedback, this statement of the second law breaks down and positive work can be produced, as exemplified by Maxwell's and Szilard's thought experiments \cite{par15,lut15}. For  Markovian feedback protocols a refined version  of the second law reads $\dot{\mathcal{W}}_{\text{ext}}\leq\kb T_0 \dot{\mathcal{I}}_\text{flow}^{\text{mar}}$,
where $T_0$ is the bath temperature and $\dot{\mathcal{I}}_{\text{flow}}^\text{mar}$ the information flow to the detector, defined as the time variation of the mutual information between a variable and its measured value~\cite{tou00,sagawa2009minimal,sag12}. This inequality has been experimentally verified with colloidal particles~\cite{toyabe2010experimental,roldan2014universal} and single electrons~\cite{koski2014experimental,koski2014prl}. When  the information rate $\dot{\mathcal{I}}_{\text{flow}}^\text{mar}$ is positive, more work can be extracted from the system than permitted by the usual second law of thermodynamics.
\begin{figure}	\includegraphics[width=0.95\linewidth]{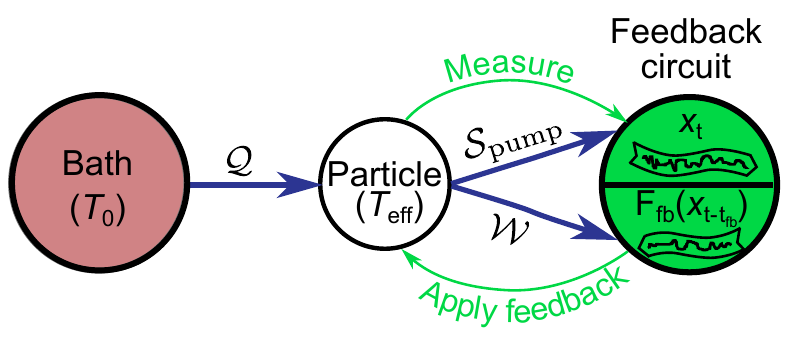}
	\caption{\textbf{Non-Markovian delayed feedback control.} The system consists of a particle (white) in contact with a single heat bath (red) and coupled to a feedback reservoir (green). The feedback loop continuously detects the position $x_t$ of the particle and applies a delayed feedback force $F_{\text{fb}}(x_{t-t_\text{fb}})$ proportional to the position at time $t-t_\text{fb}$. Depending on the time delay $t_\text{fb}$, the feedback cools or heats the system to a steady state effective temperature $T_{\text{eff}}$. The blue arrows in the figure correspond to cooling, when the feedback control pumps the entropy $\mathcal{S}_\text{pump}$ out of the system. Energy flows as heat $\mathcal{Q}$ from the bath to the particle and is extracted as work $\mathcal{W}$ by the feedback circuit (blue arrows). In the case of heating, the energy and entropy flows change direction. In both regimes the generalized second law holds in the form, $\dot{\mathcal{W}}_{\text{ext}} = -\dot{\mathcal{W}} \leq\kb T_0 \dot{\mathcal{S}}_{\text{pump}}$.
	\label{fig_baths}} 
\end{figure}

The fact that a control signal cannot be applied instantaneously implies that feedback circuits inevitably exhibit memory effects and are thus non-Markovian. The Markovian approximation is only valid when the delay, i.e. the time between measurement and feedback, is much smaller than  the typical timescales of the system. Delayed feedback is widespread in many areas, from chaotic systems to biology~\cite{astrom2010,jus97,boc05,arbib2003,hu13}, emphasizing the crucial need to expand the second law to account for finite memory. However, such generalization is nontrivial. Because of the non-Markovian nature of the feedback, the conventional approach of stochastic thermodynamics \cite{sei12} cannot be applied and the usual condition of local detailed balance does not hold. As a result,  new contributions to the nonequilibrium entropy production occur, leading to the extended second law for continuous, non-Markovian feedback,
$\dot{\mathcal{W}}_{\text{ext}}\leq\kb T_0 \dot{\mathcal{S}}_{\text{pump}}$, 
where $\dot{\mathcal{S}}_{\text{pump}}$ is the entropy pumping rate which  incorporates the effect of the time delay \cite{mun14,rosinberg2015stochastic,ros17,horowitz2014}. Since $\dot{\mathcal{S}}_{\text{pump}}\leq \dot{\mathcal{I}}_{\text{flow}}^\text{mar}$, this is the tightest second-law inequality to date~\cite{rosinberg2015stochastic}. Despite  the omnipresence of delay in feedback processes, a dedicated experimental investigation of this non-Markovian generalization of the second-law inequality is still lacking.

Here, we report the experimental study of the thermodynamics of continuous, non-Markovian feedback control applied to the underdamped center-of-mass (CM) motion of a levitated microsphere \cite{grass2016optical}. Levitated particles are an ideal experimental platform to explore thermodynamics in small systems~\cite{Rondin,Ricci,Gieseler18,Millen14,Gieseler14,Li18}. We confirm the validity of the generalized second law for time delays spanning two decades and observe the breakdown of the high-quality-factor (high Q) approximation~\cite{rosinberg2015stochastic}.  We establish that the efficiency of the feedback is enhanced when non-Markovian effects are included. We further explore the relation between Markovian and non-Markovian feedback by analyzing how the delay affects the correlations between measurement outcome and  velocity of the particle. We finally explore the limitations of feedback cooling and the saturation of the effective CM temperature of the system above the bath temperature for very large delay.

\begin{figure}	\includegraphics[width=0.99\linewidth]{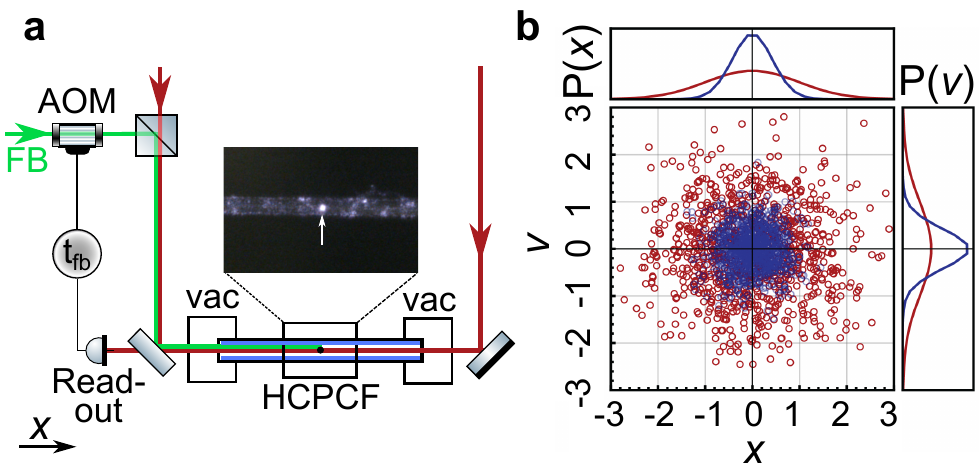}
	\caption{\textbf{Experimental setup}. \textbf{a)} Two counterpropagating laser beams (red) are coupled into a hollow-core photonic crystal fiber (HCPCF). They form a standing wave that allows for trapping a silica microparticle at an anti-node (inset, white arrow). The fiber ends are placed inside vacuum chambers (vac) to control the pressure inside the HCPCF. The particle position along the fiber axis is detected with an interferometric readout. The feedback loop is implemented delaying the detected signal by a time $t_\text{fb}$. To apply the feedback force, a feedback laser (green) is modulated with the delayed signal via an acousto-optic modulator (AOM). \textbf{b)} Phase space distribution $P(x,v)$ of the microparticle derived from the position measurements at $T_{\text{eff}}=T_0$ (red circles) and $T_{\text{eff}}<T_0$ (blue circles). The respective marginals for the  distributions of position and velocity are shown on the top and right panels.\label{fig_Setup}}  
\end{figure}

\vspace{.1in}
\noindent {\bf \large{\textsf{Results}}}\\
\textbf{Generalized second law.}
We consider a harmonic oscillator in contact with a heat bath at temperature $T_0$, and subjected to a delayed feedback control that acts as an information reservoir (Fig.~\ref{fig_baths}). Delayed feedback control means here that we acquire information about the oscillator position $x_t$ at time $t$ and apply a force $F_{\text{fb}}(t) \propto x_{t-t_{\text{fb}}}$ to manipulate its motion based on the position measured at time $t-t_{\text{fb}}$. The stochastic dynamics of the oscillator is  governed by the underdamped Langevin equation \cite{ris89},
\begin{equation}\label{eq_motion}
\ddot{x}_t+ \Gamma_0 \dot{x}_t+ \Omega_0^2 x_t- g \Gamma_0 \Omega_0 x_{t-t_{\text{fb}}}=\sqrt{\frac{2\Gamma_0 k_BT_0}{m}}\xi_t
\end{equation}  
with $m$ the particle mass, $\Omega_0$ its natural frequency and $\Gamma_0$ the damping coefficient. The quantity $\xi_t$ is a centered Gaussian white noise with $\braket{\xi(t)\xi(t')}=\delta(t-t')$. The linear feedback is applied via the force  $F_{\text{fb}}= -g (m \Gamma_0 \Omega_0) x_{t-t_{\text{fb}}}$ with feedback gain $g>0$ \cite{rosinberg2015stochastic}. The mechanical quality factor of the resonator is given by $Q_0=\Omega_0/\Gamma_0$ and the feedback damping rate by $\Gamma_{\text{fb}}=g \Gamma_0$. 
It is convenient to introduce the normalized delay $\tau=t_\text{fb} \Omega_0$ and the position of the oscillator normalized to its standard deviation in equilibrium \cite{mun14,rosinberg2015stochastic,ros17,horowitz2014}. 
The dynamics of the particle is then fully characterized by a set of dimensionless parameters $(g,Q_0,\tau)$ (Methods).  Both the harmonic and Brownian terms in Eq.~\eqref{eq_motion} are Markovian.
Memory effects enter only via the delayed feedback force $F_{\text{fb}}$.

When the feedback force acts to cool the harmonic oscillator, the extracted work $\mathcal{W}_{\text{ext}}=-\mathcal{W}$  is taken to be positive. According to the first law, the energy balance reads $\Delta U = \mathcal{W} -\mathcal{Q}$. As result, the  heat dissipated into the  heat bath  in the steady state is $\mathcal{Q}=-\mathcal{W}_{\text{ext}}$. On the other hand, the steady-state entropy balance is
\cite{mun14,rosinberg2015stochastic,ros17,horowitz2014},
\begin{equation}\label{2ndlaw}
-\frac{\dot{\mathcal{Q}}}{k_B T_0}=\frac{\dot{\mathcal{W}}_{\text{ext}}}{k_BT_0}\leq \dot{\mathcal{S}}_{\text{pump}},
\end{equation}
where $\dot{\mathcal{S}}_{\text{pump}}$ is the entropy pumping rate,  an additional contribution to the  entropy production that stems from the feedback control and depends on the delay for non-Markovian protocols. The entropy pumping rate is  computed by  coarse-graining the harmonic and feedback forces over the position variable \cite{mun14,rosinberg2015stochastic,ros17,horowitz2014} (Methods). For a harmonic potential and a linear feedback force, the velocity distribution is a Gaussian (Fig.~\ref{fig_Setup}b), $P(v)=\exp\pare{-v^2/2\sigma^2_v}/\sqrt{2\pi\sigma^2_v}$ with variance $\sigma^2_v$.  The  non-Markovian feedback control leads to a cooling of the microparticle, corresponding to negative entropy pumping and extracted work rates,  when $\sigma^2_v <1$. By contrast, heating occurs for $\sigma^2_v >1$.

\begin{figure*}[ht!]	\includegraphics[width=0.95\linewidth]{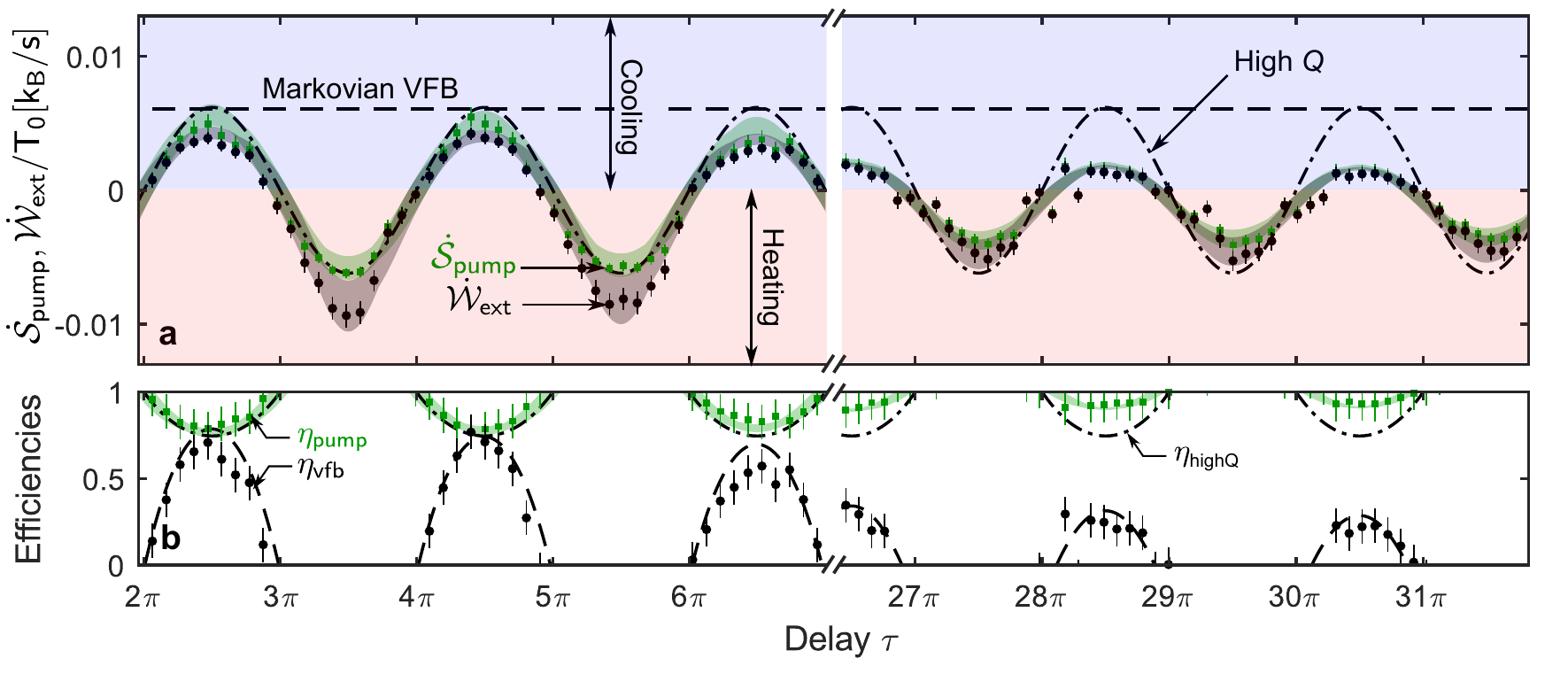}
	\caption{\textbf{Test of the generalized second law}. \textbf{a)} Experimental entropy pumping (green squares) and extracted work (black circles) rates are plotted as a function of time delay for $Q_0=55$ and $g=0.36$. The shaded area corresponds to the analytical prediction including experimental parameter drifts (Supplementary Note 8). The error bars represent statistical uncertainties. Theoretical predictions for Markovian entropy bounds to the extracted work are shown for velocity feedback ($\dot{\mathcal{S}}_{\text{vfb}}$, horizontal dashed line) and for the high-quality-factor approximation ($\dot{\mathcal{S}}_{\text{highQ}}$, dashed-dotted line). The blue (red) region corresponds to cooling (heating) of the particle motion. For all delays, the entropy  pumping is larger than the extracted work. At the same time, it represents a tighter bound to the extracted work than the Markovian predictions. \textbf{b)} 
	Feedback efficiencies plotted in the cooling regions only. Experimental pumping efficiency $\eta_{\text{pump}}=\dot{\mathcal{W}}_{\text{ext}}/(k_BT_0\dot{\mathcal{S}}_{\text{pump}})$ (green squares) of the generalized 2nd law is compared to the experimental Markovian velocity feedback efficiency $\eta_{\text{vfb}}=\dot{\mathcal{W}}_{\text{ext}}/(k_BT_0\dot{\mathcal{S}}_{\text{vfb}})$ (black circles) where $\dot{\mathcal{S}}_{\text{vfb}}=g/Q_0$. Theoretical Markovian predictions for $\eta_{\text{vfb}}$ (dashed) and $\eta_{\text{highQ}}=\dot{\mathcal{W}}_{\text{ext}}/(k_BT_0\dot{\mathcal{S}}_{\text{highQ}})$ (dashed-dotted) with $\dot{\mathcal{S}}_{\text{highQ}}=(g/Q_0)\sin\tau$ are shown for comparison. Neglecting the non-Markovian behavior leads to a strongly underestimated efficiency of the feedback process.
		\label{2nd_law_7}} 
\end{figure*}

The validity of inequality~(\ref{2ndlaw}) can be assessed by comparing the pumping entropy to Markovian bounds. The usual Markovian velocity feedback (VFB) cooling \cite{kim2004entropy,kim07}, with a feedback force proportional to the instantaneous velocity of the particle, $F_{\text{fb}}\propto -v$, is recovered in the limit of $Q_0\gg \tau$ and for $\tau=\pi/2 +2\pi n$, with $n$ an integer. The entropy pumping rate corresponds in this case to $\dot{\mathcal{S}}_{\text{vfb}}=g/Q_0$~\cite{mun13}, which we identify as the Markovian information flow  $\dot{\mathcal{I}}_\text{flow}^\text{mar}$ (Methods).  The high-quality-factor approximation ($Q_0\gg1$) allows one to map the non-Markovian dynamics to an effectively Markovian Langevin equation, because the motion of the oscillator is essentially coherent on that timescale~\cite{rosinberg2015stochastic}.
The effect of the feedback is then incorporated in a modified damping $\Gamma'=\Gamma_0(1+g\sin\tau)$ and mechanical frequency $\Omega'^2=\Omega_0^2[1-({\Gamma_{\text{fb}}}/{\Omega_0})\cos\tau]$ of the resonator (Supplementary Note 2). The Markovian second law remains valid with these modified parameters. In the high Q approximation, the entropy pumping is given by $\dot{\mathcal{S}}_\text{highQ}=(g/Q_0)\sin\tau$~\cite{rosinberg2015stochastic}. The approximation is expected to break down for larger $\tau$, when the Brownian force noise leads to dephasing between the oscillator motion and the feedback signal. This regime can only be correctly described using the generalized second law (\ref{2ndlaw}).   

 \noindent \textbf{Experimental setup and results.}
In our experiment, we use an optically levitated microparticle to implement the dynamics of Eq.~\eqref{eq_motion}, that holds for any harmonic system with linear feedback control. A standing wave is formed by two counterpropagating laser beams ($\lambda=1064$ nm) inside a hollow-core photonic crystal fiber (HCPCF) (Fig.~\ref{fig_Setup}a and Methods) \cite{grass2016optical}. A silica microsphere (969 nm diameter) is trapped at an intensity maximum of the standing wave. The amplitude of the particle motion is sufficiently small to allow for a harmonic approximation of the potential with frequency $\Omega_0/2\pi=404$~kHz (Supplementary Note 7). The damping coefficient $\Gamma_0$ as well as the bath temperature $T_0=293$~K  are determined by the surrounding gas. In our setup, the linear dependence of the damping coefficient $\Gamma_0$ on the environmental pressure allows simple and systematic tuning of this parameter along with the mechanical quality factor $Q_0=\Omega_0/\Gamma_0$. The particle motion along the $x$-axis is detected by interferometric readout of the light scattered by the particle \cite{grass2016optical}. The signal is fed into a delay line that is digitally implemented and the output signal serves to control the power of a feedback laser. This laser exerts a radiation pressure force in one direction, accelerating the microparticle proportional to the delayed particle position. The overall amplification of the signal sets the proportionality constant, which is given by $g \Omega_0 \Gamma_0$, where the gain in our experiment is $g=0.36$ (Supplementary Note 5). The whole feedback circuit has a minimal delay of $t_{\text{fb}}=2.6\mathrm{\mu s}$, i.e., $ \tau=2.04\pi$.

We first test the extended second-law  \eqref{2ndlaw} by varying the delay over two decades. Figure~3a demonstrates  the validity of the non-Markovian inequality \eqref{2ndlaw} over  all relevant timescales.
The non-Markovian entropy pumping rate $\dot{\mathcal{S}}_{\text{pump}}$ (green) is a much more precise upper bound to the  extracted work rate  (black) than the Markovian pumping rate $\dot{\mathcal{S}}_{\text{vfb}}$ (horizontal dashed line). In particular, the Markovian result fails to capture the oscillations  of the extracted work rate, as well as the heating phases induced by the delay. The high Q approximation  (dashed-dotted line) correctly describes the oscillatory behavior  of $\dot{\mathcal{S}}_{\text{pump}}$ for short delays. Yet, it does not account for the oscillation decrease induced by the Brownian force noise for long delays. We already observe significant deviations for a delay of only three oscillation periods with a mechanical quality factor of $Q_0=55$. We have also verified the second law~(\ref{2ndlaw}) by varying the dissipation via $Q_0$ (Supplementary Note 9).

The generalized second law \eqref{2ndlaw} is crucial to properly estimate the performance of the feedback cooling. In analogy to heat engines and refrigerators, one may define the efficiency of work extraction $\eta_\text{pump}= \dot{\mathcal{W}}_{\text{ext}}/(k_BT_0\dot{\mathcal{S}}_{\text{pump}})$, which characterizes the conversion of information into extracted work~\cite{cao2009}. As shown in Fig.~\ref{2nd_law_7}b, the corresponding Markovian efficiencies for velocity feedback ($\eta_{\text{vfb}}$) and for the high Q approximation ($\eta_{\text{highQ}}$) vastly underestimate the feedback efficiency. We note that pumping  efficiency $\eta_\text{pump}$ and cooling power $\dot{\mathcal{W}}_{\text{ext}}$ exhibit a trade-off similar to that of heat engines: one is maximal when the other is minimal, and vice versa. By contrast, the  velocity feedback efficiency $\eta_\text{vfb}$ exhibits an opposite dependence on $\tau$. 

\begin{figure}[t!]\includegraphics[width=0.9\linewidth]{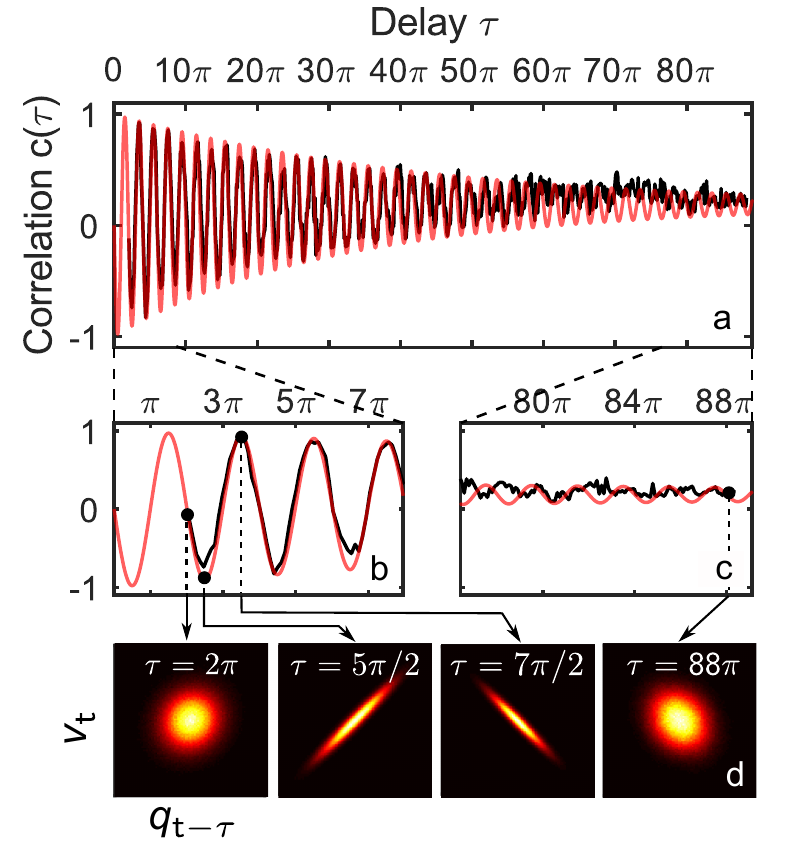}
	\caption{\textbf{Correlation and colored noise for long delays}. \textbf{a)} Experimental (black) and theoretical (red) correlation functions of the microparticle CM motion as a function of $\tau$ for $g=0.36$. \textbf{b,c)} Zoom in for short (b) and long (c) delays. \textbf{d)} Experimental probability distribution $P(q_{t-\tau},v_t)$ for different values of time delay in panels b) and c) indicated by black arrows. The system is still correlated for very long delays ($\tau=88\pi$) due to resonant driving by colored force noise created by the feedback loop.
	} 
	\label{fig_corr}
\end{figure}

We may gain physical insight on the breakdown of the standard second law seen in Fig.~\ref{2nd_law_7}  by analyzing the correlations between particle velocity  feedback force, as well as the effective temperature of the system.
Cooling is efficient when the feedback force counteracts the motion of the oscillator, in other words, when the velocity $v_t$ of the oscillator and the feedback force ($F \propto x_{t-\tau}$) are anticorrelated. Heating  occurs when they are correlated.  Figure~\ref{fig_corr} shows the correlation function between the two quantities, $c(\tau)=1/(\sigma_q\sigma_v)\int\int y_tv_tP(y_t,v_t)\text{d}y_t\text{d}v_t$ with $y_t=q_{t-\tau}$  (Supplementary Note 4). The delay $\tau$ has two effects. First, it changes the phase between the mechanical system and the feedback signal deterministically, resulting in the oscillatory behaviour of the correlations, and thus the difference between heating and cooling. Second, it allows for stochastic dephasing of the mechanical motion with respect to the feedback signal, which translates into a reduction of the correlations for increasing delay. These correlations do not vanish, however, but asymptotically approach a finite value. For long delays, the oscillator thermalizes due to the damping. The action of the feedback circuit can then be seen as an independent force noise with the spectrum of a white-noise driven harmonic oscillator. The positive correlations occur as a result of the resonant driving of the mechanical motion by the feedback signal.

\begin{figure}[b!]\includegraphics[width=0.99\linewidth]{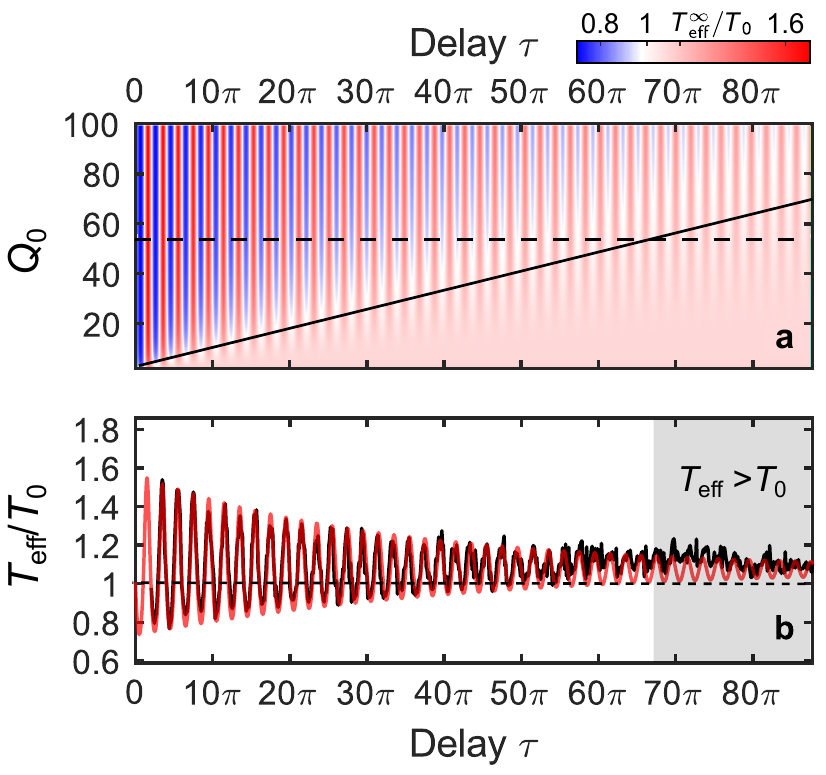}
	\caption{\textbf{Steady-state effective temperature}. \textbf{a)} Calculated 2D color map of the ratio $T_{\text{eff}}/T_0$ of the microparticle motion as a function of $\tau$ and $Q_0$ for $g=0.36$. The black solid line represents the border beyond which cooling is no longer possible. Note that this border is gain dependent. \textbf{b)} Experimental (black) and calculated (red) normalized effective temperature for $Q_0=55$, corresponding to the dashed horizontal line shown  in panel a). The grey shaded area shows the region with $T_{\text{eff}}>T_0$ for very long delays.} 
	\label{fig_temp}
\end{figure}

Figure~\ref{fig_temp} displays the ratio of the effective steady-state temperature, $T_{\text{eff}}=T_0\sigma_q^2$ \cite{rosinberg2015stochastic}, and the bath temperature $T_0$. Figure~\ref{fig_temp}a clearly shows  how the action of the feedback is reduced when  the mechanical quality factor $Q_0$ is decreased or  the time delay $\tau$ increased. For all values of $Q_0$, there is  a certain delay $\tau$, beyond which cooling is no longer possible (black line). 
For even longer delays, the effective temperature reaches a constant value, $T_{\text{eff}}^{\infty}/T_0\approx 1+g^2/2$, that is independent of $Q_0$ for weak coupling $g\ll Q_0$ (Supplementary Note 3). This is in line with the second law that predicts an asymptotic negative work extraction, $\mathcal{W}_\text{ext}^{\infty} \approx -g^2/(2Q_0)$ for very long delays. Figure~\ref{fig_temp}b provides a cut for constant $Q_0=55$ through Fig.~\ref{fig_temp}a. The grey-shaded area on the right shows the region where $T_\text{eff}>T_0$ for our gain $g=0.36$. Note that this region can be reduced by decreasing the feedback gain. Excellent agreement between theory (red) and data (black) is observed. 

\vspace{.1in}
\noindent {\bf \large{\textsf{Discussion}}}\\
In summary, we have performed an extensive experimental study of the second law of thermodynamics in the presence of continuous non-Markovian  feedback. Our results constitute an important step towards bridging theoretical developments in stochastic information thermodynamics and more technical applications like continuous feedback. Possible generalizations include the consideration of measurement noise \cite{mun13} and  the extension to nonlinear potentials and nonlinear feedback (e.g. parametric cooling of levitated nanoparticles \cite{gieseler12}). Nonlinear delayed feedback control has thus already proven to be a more robust and effective method for synchronization compared to its linear counterpart~\cite{popovych2005effective}. Another avenue of future research is the use of optimal control via Kalman-Bucy filters~\cite{horowitz2014}. Intuitively, one might hope that elaborate filtering methods, like Kalman filters, may overcome the impact of delay that we observe in our simple feedback scenario by an optimal prediction of the instantaneous velocity. However, while this will help to reduce the effect of measurement noise,  the effect of stochastic Brownian force noise that occurs during the delay is fundamentally unpredictable. We therefore anticipate no improvement compared to the long delay results presented in our work.

\vspace{.25in}

\noindent {\bf \large{\textsf{Methods}}}\\
{\small
\textbf{Normalized form of the equation of motion.}
After time contraction $t\rightarrow t\Omega_0$ and position normalization $x\rightarrow q=x/x_{\text{th}}$ with $\xth=\sqrt{\kb T_0/{m\om^2}}$ the thermal root mean square amplitude, the Langevin equation can be rewritten in the dimensionless form,

\begin{equation}\label{eq_motion_no}
	\ddot q_t + \frac{1}{Q_0}\dot q_t + q_t - \frac{g}{Q_0}q_{t-\tau} = \sqrt{\frac{2}{Q_0}}\xi_t.
\end{equation}
with $Q_0=\om/\Gamma_0$ the quality factor, $g=\gafb/\Gamma_0$ the feedback gain, $\tau=t_{\text{fb}}\Omega_0$ the normalized delay and $\xi_t$ representing a centered Gaussian white noise force with $\braket{\xi(t)\xi(t')}=\delta(t-t')$. 
\bigskip

\noindent\textbf{Experimental setup.} 
The particle is trapped in an intensity maximum of the standing wave and oscillates in a harmonic trap. For a laser power of 400 mW in each trapping beam, the mechanical frequency is $\Omega_0/2\pi= 404$ kHz. The particle CM motion is recorded using a balanced photodiode and roughly 10\% of the light transmitted through the HCPCF and that scattered by the particle. The feedback control is implemented via radiation pressure of a second laser beam which is orthogonally polarized and frequency shifted with respect to the trapping laser to avoid interference effects. The feedback force $\fb$ is realized in the following steps. The read-out signal of the CM motion is bandpass filtered (bandwith: 600 kHz, center frequency: $\om/2\pi$) to get rid of technical noise in the detector  signal. Then, the signal can be amplified and delayed by an arbitrary time $t_{\text{fb}}$ with a field programmable gate array (FPGA). This signal is then used as modulation input for the AOM. A more detailed description of the experimental setup and the feedback control can be found in Supplementary Note 6 and in Ref.~\cite{grass2016optical}. 
\bigskip

\noindent\textbf{Thermodynamic quantities.} 
 The non-Markovian entropy pumping rate $\dot{\mathcal{S}}_{\text{pump}}$ can be computed    by coarse-graining the harmonic (h) and feedback (fb) forces over the position variable: $\dot{\mathcal{S}}_{\text{pump}}=-\int\diff v [\overline{F_\text{fb}(v)}+\overline{F_\text{h}(v)}]\partial_v P(v)$, where $P(v)$ is  the velocity distribution and $\overline{F_\text{i}(v)} = \int dx \overline{F_\text{i}(x,v)} P(x|v)$ the corresponding coarse-grained forces \cite{mun14,rosinberg2015stochastic,ros17,horowitz2014}. On the other hand, the Markovian information flow $\dot{\mathcal{I}}_{\text{flow}}^\text{mar}$ in the case of VFB is given by the time variation of the mutual information between the velocity $v$ and the measured value of the velocity $y$: $\dot{\mathcal{I}}_{\text{flow}}^\text{mar}=\int \text{d}v\text{d}y\partial_v J(v)\text{ln}[P(v,y)/P(v)P(y)]$ with   the velocity probability current $J(v)$ \cite{rosinberg2015stochastic,ros17,horowitz2014}. This quantity diverges for error-free feedback \cite{mun13}. We thus identify the Markovian entropy bound with the Markovian limit of $\dot{\mathcal{S}}_{\text{pump}}\rightarrow \dot{\mathcal{S}}_{\text{vfb}}=g/Q_0$ \cite{rosinberg2015stochastic,ros17,horowitz2014}.
 For a harmonic potential and a linear feedback force, the velocity distribution is a Gaussian, $P(v)=\exp\pare{-v^2/2\sigma^2_v}/\sqrt{2\pi\sigma^2_v}$, with variance $\sigma^2_v$.  The entropy pumping rate is then explicitly   $\dot{\mathcal{S}}_{\text{pump}}={(1-\sigma^2_v)}/(Q_0\sigma^2_v)$ and the work extraction rate reads $\dot{\mathcal{W}}_{\text{ext}}/(k_B T_0)=(1-\sigma^2_v)/{Q_0}$ (Supplementary Information). We experimentally obtain the velocity variance  needed to verify the generalized second law  as follows: for a given time delay $\tau$, a position time trace $x(t)$ is recorded, filtered with a bandwidth of $3\Gamma_0$ via post processing, and normalized with the standard deviation of the feedback beam turned off, to find the normalized position $q(t)$. The velocity $v=\dot{q}$ is then numerically calculated with finite difference approximation and the variance $\sigma_v^2$ computed.
\subsection{Data availability}
The datasets generated in the current study are available from MD on reasonable request.

\bibliographystyle{naturemag}

\section*{Acknowledgments}
We thank Markus Aspelmeyer for valuable discussions and support, and James Millen for helpful comments. N. K. acknowledges support from the Austrian Science Fund (FWF): Y 952-N36, START. D.G. acknowledges support
through the doctoral school CoQuS (Project W1210). We further acknowledge financial support from the German Science Foundation (DFG) under project FOR 2724.

\section*{Author contributions}
M.D. and D.G. carried out experiments and analysed data. N.K. and E.L conceived the experiment. J.J.A. and E.L. supported theoretical aspects. All authors discussed the results and contributed to the interpretation of the data and writing of the paper.  

\section*{Competing Interests}
The authors declare no competing interests.

\end{document}